\documentclass[aps,pra,showpacs, twocolumn]{revtex4}
\usepackage{graphicx}
\usepackage{bm}% bold math
\usepackage{amsmath}% equation formating

\providecommand{\abs}[1]{\left\lvert#1\right\rvert}
\providecommand{\norm}[1]{\lVert #1 \rVert}

\providecommand{\bra}[1]{\langle #1 \rvert}
\providecommand{\ket}[1]{\lvert #1 \rangle}
\providecommand{\braket}[2]{\langle #1 \rvert #2 \rangle}
\providecommand{\ketbra}[2]{\lvert  #1\rangle \langle #2 \rvert}
\providecommand{\be}{\begin{equation}}
\providecommand{\ee}{\end{equation}}
\providecommand{\ba}{\begin{eqnarray}}
\providecommand{\ea}{\end{eqnarray}}
\newcommand{\ud}{\mathrm{d}}
\newcommand{\I}{\mathrm{i}}
\usepackage{graphicx}

\providecommand{\abs}[1]{\left\lvert#1\right\rvert}
\providecommand{\norm}[1]{\lVert #1 \rVert}

\providecommand{\bra}[1]{\langle #1 \rvert}
\providecommand{\ket}[1]{\lvert #1 \rangle}
\providecommand{\braket}[2]{\langle #1 \rvert #2 \rangle}
\providecommand{\ketbra}[2]{\lvert  #1\rangle \langle #2 \rvert}
\providecommand{\be}{\begin{equation}}
\providecommand{\ee}{\end{equation}}
\providecommand{\ba}{\begin{eqnarray}}
\providecommand{\ea}{\end{eqnarray}}

\newcommand{\oper}[1]{\bm{\mathsf{#1}}}
\usepackage{mathbbol}

\global\long\def\braket#1#2{}

\global\long\def\ketbra#1#2{}

\global\long\def\ud{}

\global\long\def\eigvbar#1#2{}

\global\long\def\I{\mathrm{{i}}}

\global\long\def\acco#1{\left\{  #1\right.}

\global\long\def\abs#1{}

\global\long\def\norm#1{}

\begin{document}

\title{Continuous discretization of infinite dimensional Hilbert spaces}
 
\author{P. Vernaz-Gris$^1$, A. Ketterer$^1$, A. Keller$^2$, S. P. Walborn$^3$, T. Coudreau$^1$ and P. Milman$^{1,}$\footnote{ Corresponding author: perola.milman@univ-paris-diderot.fr}}

\affiliation{$^{1}$Laboratoire Mat\'eriaux et Ph\'enom\`enes Quantiques, Universit\'e Paris Diderot, CNRS UMR 7162, 75013, Paris, France}
\affiliation{$^{2}$Univ. Paris-Sud 11, Institut de Sciences Mol\'eculaires d'Orsay (CNRS), B\^{a}timent 350--Campus d'Orsay, 91405 Orsay Cedex, France}
\affiliation{$^{3}$ Instituto de F\'{\i}sica, Universidade Federal do Rio de Janeiro. Caixa Postal 68528, 21941-972 Rio de Janeiro, RJ, Brazil}

%\date{\today}
 
\begin{abstract}
In quantum theory,  observables with a continuous spectrum are known to be fundamentally different from those with a discrete and finite spectrum. While some fundamental tests and applications of quantum mechanics originally formulated for discrete variables have been translated to continuous ones, this is not the case in general. For instance, despite their importance, no experimental demonstration of nonlocality exists in the continuous variables regime.
Attempts to bridge this gap and put continuous variables on a closer footing to discrete ones used dichotomization. However, this approach considers only discrete properties of the continuum, and its infinitesimal properties are not fully exploited. Here we show that it is possible to manipulate, detect and classify continuous variable states using observables with a continuous spectrum revealing properties and symmetries which are analogous to finite discrete systems.  Our approach leads to an operational way to define and adapt,  to arbitrary continuous quantum systems, quantum protocols and algorithms typical to discrete systems. 
\end{abstract}
\vskip2pc 
 
\maketitle

Continuous observables have a continuous spectrum, that can be limited or unbounded. In both cases, continuous observables allow for a representation in terms of an infinite set of discrete states. Observables often referred to as ``discrete" have a discrete spectrum, and can be represented by a  finite set of discrete states. Examples of continuous observables  are position and momentum, the electromagnetic  field's quadrature, or the angular distribution of a confined particle. Examples of discrete observables with a finite spectrum  are the spin of  a particle or the polarization of a photon. Continuous observables are  directly related to their infinite discrete representation  by a Fourier conjugation relation, while finite discrete observables are described by a $SU(d)$ symmetry group, where $d$ is the dimension of the corresponding Hilbert space. Quantum continuous systems and finite discrete ones are thus clearly fundamentally different, and consequently, it is still unknown how they compare in what concerns, for instance, their usefulness in fundamental tests and applications of quantum mechanics. It is usually thought that each type of observables possesses its own set of advantages and drawbacks. 

In the field of quantum information, the use of continuous variables (CV) often handles gaussian states, for which necessary and sufficient conditions for detecting entanglement exist \cite{SIMON, DUAN}. Some quantum information tasks, as teleportation \cite{TELEPORTATION}, have been realized unconditionally, albeit with a limited fidelity \cite{TELEPORTATIONEXP}. Fundamental tests of quantum mechanics as, for instance, Bell-type inequalities \cite{BELL}, can be formulated in terms of variances of distributions, that can be gaussian or non-gaussian (see Ref. \cite{REID} for some examples). However, to date, no experimental demonstration of  non--locality of quantum mechanics has been done with CV. Furthermore, the field of quantum computing with CV still lacks operational solutions for defining, for instance, quantum logic gates and algorithms applying to continuous systems  \cite{RMPBrau, Wedbrook}. 

When one deals with discrete observables, different types of problems and advantages appear. A clear advantage of discrete systems is the ability to define quantum algorithms and protocols outperforming  their classical analogs \cite{SHOR, GROVER}. Also, correlations between incompatible observables can reveal intrinsically quantum behavior in a simple and operational way, as in the Clauser--Horn--Shimony--Holt (CHSH) formulation of Bell--type inequalities \cite{CHSH}. However, discrete variables usually require manipulation and measurements in the single particle scale,  which are at the origin of experimental difficulties that haven't been systematically overcome yet. 

 An usual solution to bridge the worlds of discrete and continuous variables is using dichotomization, or digitalization techniques. It consists in manipulating and measuring continuous variables states using  observables with a discrete spectrum, such that the space of states becomes classifiable according to this observable's eigensystem. One example of observable leading to dichotomization is parity,  that was used in the non locality test proposed by W\'odkievicz and Banazsek \cite{BANA, pseudo, Saleh}.  

In the present contribution we show that it is possible to define  observables with a continuous spectrum enabling an operational manipulation and detection of continuous variables systems, analogously to finite discrete systems. As a consequence, we show that it is possible to generalize to continuous systems, algorithms and protocols originally conceived for finite discrete systems. This is done by simply replacing the usual $SU(d)$ operators by the ones defined in our formulation, that present several analogies with the $SU(d)$ group. 

This article is organized as follows: we first recall the formalism of modular variables, that is an essential tool to derive our main results. Using this formalism, we are able to  define a continuum of discrete subspaces of arbitrary dimension $d$. In each subspace, $SU(d)$ generators can be defined, and considering the infinite sum of all the subspaces results in a continuum of $SU(d)$--type operators. We discuss some properties of such operators and, in particular, their differences with dichotomization using parity operators. Finally, we illustrate the application of our results introducing Bell--type inequalities and entangling gates for continuous variable systems.
 
The formalism  of modular variables was introduced by Y. Aharanov, H. Pendleton and  A. Petersen in the 60's  \cite{AHARANOV}. It explicits the  importance  of defining some ``discreteness" in the continuum in order to identify its quantum properties. It is usually applied to systems with some pre-defined length scale and periodicity, as for instance, multiple slit experiments, where each slit is separated from each other by a length $l$. Aharanov and co--workers found convenient to define position ($\hat x$) and momentum ($\hat p$) operators as follows:
\begin{eqnarray}\label{modvar}
&&\hat x=\hat {\bar x}+\hat n l \nonumber \\
&&\hat p=\hat {\bar p}+\hat m \left ( \frac{\hbar}{l} \right ),
\end{eqnarray}
where $\hat {\bar x}=\hat x $ (mod $ l)$,  $\hat {\bar p}=\hat p$ (mod $   1/l  )$.  In Ref.  \cite{AHARANOV}, the non--integer, or modular part of the momentum, $\bar p$,  appears as a consequence of the Aharanov-Bohm effect  \cite{AHARANOV2} when solenoids are placed between two slits. Recently, it was shown that variance based entanglement witnesses can be constructed using this formalism \cite{RUSSOS}, a result that was extended to entropy based witnesses and tested in a multi-slit photon pair experiment  \cite{STEVE} . These results evidence that  the modular variable formalism is useful in experimental contexts where periodicity and discretization appear naturally, in the form of periodic spatial regions, or slits. However, nothing prevents it to be generalized  to any type of experimental context, including, for instance, a freely  propagating wave--packet. Here we use the formalism of modular variables  to show that it is possible to define, in an arbitrary experimental context, $SU(d)$--type  generators using genuinely continuous observables. By such, we mean that no dichotomization or digitalization is performed, and no {\it a priori} scale, related to some physical property of the system,  should be necessarily defined, as is the case in the aforementioned experiments. Our results are relevant from the fundamental point of view, since they extract properties that were thought to be particular to discrete or discretized systems using genuinely continuous observables only. From a more practical point of view, it provides an operational way to extend to continuous variable systems quantum protocols originally conceived  for discrete systems.

We start by defining dimensionless operators $\hat \theta= \hat x \left (2\pi/l\right )$ and $\hat k=\hat p\left ( l/\hbar \right )$, so that the eigenvalues of $\hat{\bar{\theta}} \in [0,2\pi[$,
and the eigenvalues of $\hat{\bar{k}} \in [0,1[$. In \cite{SI}, we review some properties of modular variables that are useful for deriving our main result.  In particular, we have that $[\bar \theta, \bar k]= 0$, allowing the definition of a modular complete basis $\ket{\left \{\bar \theta, \bar k \right \}}$, a result that is connected to the Zak representation \cite{Zak1, Zak2}, that was used in the quantum information context in Ref. \cite{Englert}. An arbitrary quantum state $\ket{\psi}=\int d\theta g(\theta) \ket{\theta}$ can be expressed in the modular basis as 
\begin{equation}\label{state}
\ket{\psi}=\int_{0}^{1}\int_{0}^{2\pi}{\rm d}\bar{k}{\rm d}\bar{\theta}\tilde g(\bar \theta, \bar k)\ket{\left \{\bar{\theta},\bar{k}\right \}},
 \end{equation} 
where $\tilde g(\bar \theta, \bar k)$ is a normalized complex function.

\begin{figure}
\centerline{\includegraphics[width=.75\columnwidth]{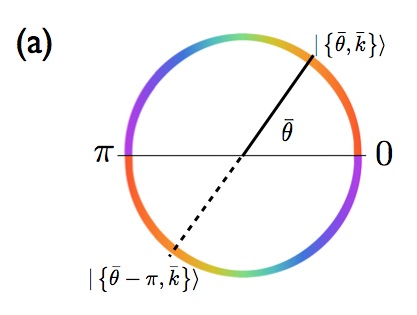}}
\centerline{\includegraphics[width=.8\columnwidth]{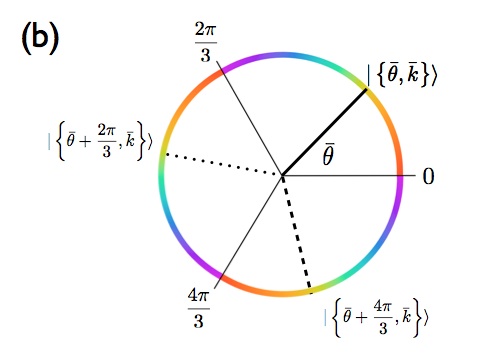}}
\caption{ Space of states for a given constant value of $\bar k$ and varying $\bar \theta$: (a) For every fixed value of $\bar k$, the space of $\bar \theta$ variable forms a ring. $SU(2)$--type subspaces are defined by splitting this ring into two regions,  such that $0 \leq \bar \theta < \pi$ in one region and  $\pi \leq \bar \theta < 2\pi$ in the other. Two states differing by $\pi$ (associated to the same color in the figure)  form a two dimensional  subspace, as indicated by the dashed and continuous lines in the Figure. In this subspace, color dependent ($\bar \theta, \bar k$ dependent) $SU(2)$ operators can be defined. The $\hat \Gamma_{\alpha}^{(d)}$ operators are an infinite sum of such color dependent $SU(2)$ operators with a weight function satisfying the periodicity constraint, since the continuity of the operators must be ensured when intervals are connected. (b)  Same as (a) for $d=3$. In this case, the  $ [0,2\pi[ $ interval is divided in three continuous zones. States differing by $2\pi/3$  belong to the same subspace (same color) and the continuity of the operators is also imposed. 
 \label{fig1}}
\end{figure}

We now  move to our main result, showing how to define observables with a continuous spectrum and presenting  some $SU(d)$--type properties that are useful to manipulate and extract information from continuous variable states.  We start by discussing  in detail the case of $SU(2)$--type observables, that contains all the general principles involved in the case of arbitrary $d$. For each $\bar k$, we consider two intervals in the $\bar \theta$ space, such that $0 \leq \bar \theta < \pi$ in one interval and  $\pi \leq \bar \theta < 2\pi$ in the other.  This is pictorially illustrated in Fig. \ref{fig1}(a) where the colored ring represents the space of states with varying $\bar \theta$ and constant $\bar k$. States distant of $\pi$ in the circle are represented by the same color. We have thus a continuum of two--level systems, composed by pairs of states $\ket{\{\bar \theta, \bar k\}}, \ket{\{\bar \theta+\pi, \bar k\}}$ that have the same color. Through this picture, we can see the Hilbert space as an infinite sum of two dimensional Hilbert spaces, where $\bar \theta, \bar k$ dependent Pauli matrices can be defined \cite{Comment}: 
 \begin{eqnarray}\label{pauli} 
&&\hat \sigma_1(\bar \theta, \bar k) \equiv \ket{\{\bar \theta, \bar k \}}\bra{\{\bar \theta, \bar k\}}-\ket{\{\bar \theta+\pi, \bar k\}}\bra{\{\bar \theta+\pi, \bar k\}} \nonumber \\
&&\hat \sigma_2(\bar \theta, \bar k) \equiv \ket{\{\bar \theta, \bar k \}}\bra{\{\bar \theta+\pi, \bar k\}}+\ket{\{\bar \theta+\pi,\bar k\}}\bra{\{\bar \theta, \bar k\}} \\
 &&\hat \sigma_3(\bar \theta, \bar k) \equiv i\left (\ket{\bar \theta, \bar k \}}\bra{\{\bar \theta+\pi, \bar k\}}-\ket{\{\bar \theta+\pi,\bar k \}}\bra{\{\bar \theta, \bar k\}} \right )\nonumber 
 \end{eqnarray}

 %\begin{figure}[h]
 %\centerline{\includegraphics[width=0.75\columnwidth]{Table1.pdf}}
 %\end{figure}
 
From (\ref{pauli}), we can construct an observable with a continuous spectrum by integration over $\bar k$ and $\bar \theta$ with a judiciously chosen weight function $\zeta_{\alpha}^{(2)}(\bar \theta, \bar k)$ (see \cite{SI} for details): 
\begin{equation}\label{SU2}
\hat \Gamma_{\alpha}^{(2)}=\int_0^{\pi}{\rm d}\bar \theta \int_0^{1}{\rm d}\bar k \zeta_{\alpha}^{(2)}(\bar \theta, \bar k)\hat \sigma_{\alpha}(\bar \theta, \bar k),
 \end{equation} 
where $\alpha=1,2,3$ and $\hat \Gamma_{\alpha}^{(2)}$ is the $\alpha$-th operator  with a continuous spectrum formed by the infinite sum of $\{\bar \theta,\bar k\}$ dependent $SU(2)$--type generators. A similar procedure can be performed for arbitrary $d$, so that
\begin{equation}\label{SUd}
\hat \Gamma_{\alpha}^{(d)}=\int_0^{\frac{2\pi}{d}}{\rm d}\bar \theta \int_0^{1}{\rm d}\bar k \zeta_{\alpha}^{(d)}(\bar \theta, \bar k)\hat \gamma_{\alpha}^{(d)}(\bar \theta, \bar k),
\end{equation}
where $\zeta_{\alpha}^{(d)}(\bar \theta, \bar k)$ is the continuous weight function, $\hat \gamma_{\alpha}^{(d)}(\bar \theta, \bar k)$ are the $\alpha=1,...,d^2-1$ independent matrices leading to the generators of $SU(d)$ in the $d$ dimensional $\ket{\{\bar \theta, \bar k\}}, \ket{\{\bar \theta+2\pi/d, \bar k\}}...,\ket{\{\bar \theta+2\pi(d-1)/d, \bar k\}}$ subspace. In Fig. \ref{fig1} (b) we illustrate the $d=3$ case: the colored ring is now divided in three regions where, once again, points with the same color define the same subspace. The function  $\zeta_{\alpha}^{(d)}(\bar \theta, \bar k)$ must have several properties ensuring that operators $\hat \Gamma_{\alpha}^{(d)}$ are observables. At the same time, we will focus on  observables that can be realized with current technology in experimental systems. In order to discuss the specific  properties of functions  $\zeta_{\alpha}^{(d)}(\bar \theta, \bar k)$ we notice that all diagonal $\hat \Gamma_{\alpha}^{(d)}$ operators can be written as
\begin{equation}\label{delta}
\hat \Delta = \int_0^{2\pi}{\rm d}\bar \theta \int_0^{1}{\rm d}\bar k F(\bar \theta, \bar k)\ket{\{\bar \theta, \bar k\}}\bra{\{\bar \theta, \bar k\}},
\end{equation}
provided that $F(\bar \theta, \bar k)$ fulfills some conditions: by identifying Eqs. (\ref{SUd}) and (\ref{delta}), we show in \cite{SI} that the absolute value of $F(\bar \theta, \bar k)$  must be $2\pi/d$ periodic and continuous in the edges of each $2\pi/d$ interval, ensuring that it  is a continuous function in the whole space. An example of  function satisfying these conditions is  $F(\bar \theta, \bar k)=\cos{(\bar \theta-\bar k \pi)}$. In the case of a continuum of $SU(2)$--type operators, this leads to $\zeta_{\alpha}^{(2)}(\bar \theta, \bar k)=\cos{(\bar \theta-\bar k \pi)}$$\   \forall$$\  \alpha$ in (\ref{SU2}). In the $SU(3)$--type case, it leads to slightly more complicated   $\zeta_{\alpha}^{(3)}(\bar \theta, \bar k)$ functions that are  detailed in \cite{SI}.  Notice that such conditions lead only to the subset of {\it diagonal} operators. 
 
From the diagonal operators, one can obtain the non--diagonal ones by using combinations of displacement operations of the type  $\hat S_{\alpha}^{(d)}(\hat \theta, \hat  k)\ket{\left \{\bar \theta, \bar k\right \}}\rightarrow f_{\alpha}^{(d)}(\bar \theta, \bar  k)\ket{\left \{\bar \theta+2\pi j(\alpha)/d, \bar k\right \}} $ and their conjugate. Here, $f_{\alpha}^{(d)}(\bar \theta, \bar  k)$ are complex functions and $\hat S_{\alpha}^{(d)}(\hat \theta, \hat  k)$ are unitary transformations  depending on the dimension $d$ and on which one of the non--diagonal generators we want to create ($\alpha$). $j(\alpha)=1,...,d^2-1$ is an integer. It is clear that since the above operations are linear and valid for all basis states $\ket{\{\bar \theta, \bar k\}}$, we can identify 
\begin{equation}\label{s}
\hat S_{\alpha}^{(d)}(\hat \theta, \hat  k)\equiv  \!\!\int_0^{2\pi} \!\!\! d\bar \theta d\bar k f_{\alpha}^{(d)}(\bar \theta, \bar  k)\ket{\{\bar \theta+2\pi/d, \bar k\}}\bra{\{\bar \theta, \bar k\}}.
\end{equation}
  According to the specific physical scenario considered, operators  $\hat S_{\alpha}^{(d)}(\hat \theta, \hat  k)$ can have different forms and lead to a continuum of different operators in $SU(d)$. Considering the transverse degrees of freedom of a photon, operator (\ref{delta}) can be obtained with a Spatial Light Modulator (SLM) and operator (\ref{s}) by using the free propagation, lenses and SLMs (see \cite{SI}). By combining such operations by interferometry \cite{SAULO}, one can engineer all the $\hat \Gamma_{\alpha}^{(d)}$  operators, showing the possibility of  immediate  experimental implementation of our results in quantum optics experiments (see \cite{SI} for details).

A natural question is how the defined operators compare to dichotomization, based for instance, on  parity \cite{BANA}, described by operator $\hat \Pi_1$. An arbitrary quantum state can be expressed  in terms of parity eigenstates, that form a two dimensional basis. In Ref. \cite{pseudo}, two other parity related observables were defined, $\hat \Pi_2$ and $\hat \Pi_3$ so that together with parity they closed a $SU(2)$ algebra. Using this representation, every pure continuous quantum state can be represented by a point on the surface of a Bloch sphere, analogously to pure two--level systems. Its coordinates are given by the expected values of the $\hat \Pi_{\alpha}$ operators, $\alpha=1,2,3$ \cite{pseudo}. Every point on the surface of such a sphere is infinitely degenerate, since completely different quantum states can have the same parity or its conjugate properties. By considering non pure states as well, we move to a Bloch ball description, and  the analogy between parity and a two--level system also holds in this case. 

 \begin{figure}
 \centerline{\includegraphics[width=.8\columnwidth]{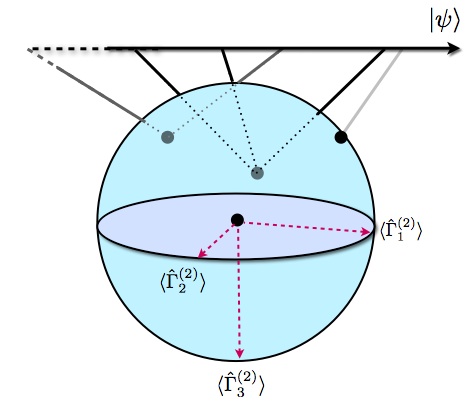}}
\caption{ Bloch ball--like representation of pure states using coordinates defined by the continuous operators $\hat \Gamma_{\alpha}^{(2)}$:. By associating to every  state a point $\langle \hat \Gamma_{1}^{(2)}\rangle, \langle \hat \Gamma_{2}^{(2)}\rangle, \langle \hat \Gamma_{3}^{(2)}\rangle $, we construct an unitary radius ball. The number of sates that are associated to a same point in the ball is not uniform, depending on the considered point. Points at the unitary radius sphere are associated to eigenstates of $\hat \Gamma_{\phi}^{(2)}$  (linear combinations of $\hat \Gamma_{\phi}^{(2)}$ in the $\phi$ direction) with eigenvalue equal to one, whereas inner points in the  ball are associated to multiple (infinite) states, as illustrated. Mixed states are inside the ball and cannot be found in the unit radius sphere 
 \label{fig2}}
\end{figure}

The present formulation is fundamentally different, even when we consider the case of $d=2$, that we now discuss in detail. As above, it is also possible to define a unit radius ball associating to every state $\ket {\Psi}$, with coordinates $\langle \hat \Gamma_{1}^{(2)} \rangle_{\psi}$, $\langle \hat \Gamma_{2}^{(2)} \rangle_{\psi}$ and $\langle \hat \Gamma_{3}^{(2)} \rangle_{\psi}$, as shown in Fig. \ref{fig2}. Operators $\hat \Gamma_{\alpha}^{(d)}$ are incompatible and  $\sum_{\alpha=1}^{d^2-1} \langle \hat \Gamma_{\alpha}^{(d)}\rangle ^2 \leq 1$ (that is also valid in the particular case of $d=2$). 

However, this ball has specific properties which are a consequence of the fact it describes continuous states.  
We start by considering the subspace of pure states. It is easy to check that the ball appearing in Fig. (\ref{fig2}) is completely filled in this case. This shows that the space of pure states that can be differentiated from each other by computing   $\langle \hat \Gamma_{\alpha}^{(2)} \rangle_{\Psi}$, $\alpha=1,2,3$ is larger than when dichotomizing, and computing $\langle \hat \Pi_{\alpha} \rangle_{\Psi}$, $\alpha=1,2,3$.  As a matter of fact, the ball depicted in Fig. \ref{fig2} can be seen as a superposition of infinite spheres with radius continuously varying from $0$ to $1$.  For each value of the radius, we have  $\bar \theta, \bar k $ dependent bi--dimensional subspaces formed by degenerate eigenstates of  $\hat \Gamma_{\phi}^{(2)}$, where $\phi$ is an arbitrary direction in the $3$ dimensional space. 

It is interesting to notice that in the present description, the number of different continuous quantum states associated to a same point in the Bloch--type sphere is not constant and depends on the point itself. 
The unitary radius sphere is formed by pure states that are the eigenstates with $\pm 1$ eigenvalue of some linear combination of $\hat \Gamma_{\alpha}^{(2)}$'s. Inner points associated to pure states are increasingly  degenerate. When considering the entire space of states, including non--pure ones, they must necessarily all be contained in the Bloch--type ball, since we dispose of three parameters to describe all the quantum states.  It is clear that the Bloch--type ball presented here is not enough to characterize a quantum state. Quantum state characterization can be improved by considering the  measurement of the expectation value of  other $\hat \Gamma_{\alpha}^{(d)}$ operators:  two states that are represented by the same set of values of $\langle \hat \Gamma_{\alpha}^{(d)} \rangle$ for some $d$ may  be distinguished when measuring $\langle \hat \Gamma_{\alpha}^{(d')} \rangle $, with $d \neq d'$.

\begin{figure}
\includegraphics[width=0.5\textwidth]{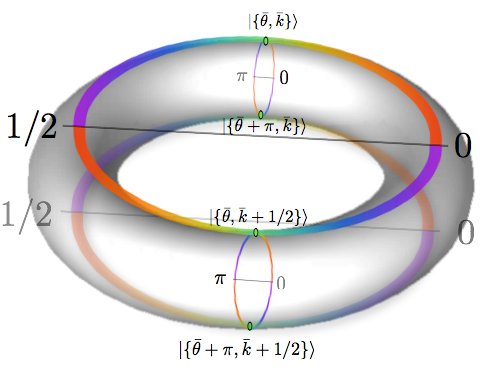}
\caption{  Geometrical picture of the Hilbert space in terms of $\ket{\{\bar \theta, \bar k\}}$ and its representation in terms of a continuum of bidimensional systems: The $\{\bar \theta, \bar k\}$ space can be represented as a torus. Fig. 1 represents a cut in this torus of fixed $\bar k$ and varying $\bar \theta$. Two of these cuts are represented in the figure, one for a given $\bar k$ and the other for $\bar k + 1/2$. Rings with constant $\bar \theta$ and varying $\bar k$ can also be defined, as shown in two examples in the Figure. Subspaces can be defined either in rings where one of the variables is fixed and the other varies, or in situations where both varies. In the latter case, $\ket{\{\bar \theta, \bar k \}}, \ket{\{\bar \theta, \bar k+1/2 \}}$ subspaces are defined, and in each subspace, subspaces $\ket{\{\bar \theta, \bar k\}}, \ket{\{\bar \theta+\pi, \bar k \}}$ and $\ket{\{ \bar \theta, \bar k+1/2 \}}, \ket{\{\bar \theta+\pi, \bar k+1/2 \}}$ are defined. 
\label{fig3}}
\end{figure}

It is important to notice that, up to now, we have considered only  displacements in the $\bar \theta$ coordinate that do not affect the $\bar k$ coordinate. In view of (\ref{modvar}), they are equivalent to changing the value of the $\hat m$ eigenstate while keeping $\bar k$ constant. It is clear that  one can realize displacements in the $\bar k$ coordinate as well (see \cite{SI}), and for each fixed $\bar \theta$, define continuous operators $\hat \Lambda_{\alpha}^{(d)}$, which are  perfectly analogous to $\Gamma_{\alpha}^{(d)}$ in the $\bar k$ coordinate. In particular, since both variables are independent we can combine both type of operators, defining, for instance, operators $\hat \Gamma_{\alpha}^{(d)}  \hat \Lambda_{\alpha'}^{(d')}$, where the values of $\alpha$ and $\alpha'$ and of $d$ and $d'$ are independent, not necessarily the same. In this more complete scenario, where a larger part of the continuous Hilbert space is being manipulated and/or measured, we can apply the same discussion presented here independently to the $\bar k$ and $\bar \theta$ variables. However, in this case, the structure of the space created by operators $\hat \Gamma_{\alpha}^{(d)}$ and $\hat \Lambda_{\alpha'}^{(d')}$ is much more complicated, even in the case $d=d'=2$: in Fig. \ref{fig3} we illustrate this by showing the torus representing the $\{\bar \theta, \bar k\}$ space and different rings corresponding to varying $\bar \theta$ for different values of $\bar k$ and vice-versa. Points with the same color can still define $d$, $d'$ or $d.d'$ subspaces, as illustrated.

We now study some applications of the introduced operators, evidencing their power to manipulate continuous variables and implement quantum protocols in continuous systems that are analogous to discrete ones. An example of such protocols are Bell-type inequalities. We define, in a bipartite system,  $\hat \Gamma_{\phi_i}^{(2)}$, $i=1,2$ as the  linear combinations of $\hat \Gamma_{\alpha}^{(2)}$ in the  $\phi_i$ direction, analogously to Pauli matrices. We can show in this case that  
\begin{equation} \label{bell} 
\langle \hat \Gamma_{\phi_1}^{(2)} \hat \Gamma_{\phi_2}^{(2)}\rangle + \langle \hat \Gamma_{\phi'_1}^{(2)} \hat \Gamma_{\phi_2}^{(2)}\rangle+\langle \hat \Gamma_{\phi_1}^{(2)} \hat \Gamma_{\phi'_2}^{(2)}\rangle-\langle \hat \Gamma_{\phi'_1}^{(2)} \hat \Gamma_{\phi'_2}^{(2)}\rangle \leq 2
\end{equation}
under the assumption of local realism. However, this inequality is violated for some entangled continuous variables states. Such states show entanglement between distributions which are ``close to eigenstates" \cite{comment2} of $ \Gamma_{\phi_i}^{(2)}$,  analogously to CHSH Bell--type inequalities \cite{CHSH, BELL}. The present results generalize those derived for continuous variables operators with a bounded spectrum  \cite{Milman07, Milman10, Borges12} to arbitrary continuous or discrete variables systems, irrespectively of their dimension or  spectral properties.  Bell--type inequalities involving correlations between $SU(d)$ operators \cite{COLLINS, QUTRITS} can also be generalized through the present formalism by using correlations between $\hat \Gamma_{\alpha}^{(d)}$ observables. 

To show how  to generalize to continuous variables quantum logic gates, circuits and algorithms, we start by considering a bipartite system, where $i=1,2$ denotes each party and for each party $d=2$ (as in Fig. \ref{fig1} (a)). Still considering  $\bar k_i$ fixed, we can thus define quantum logic gates acting for each pair $\bar \theta_1, \bar \theta_2$ by combining $\hat \sigma(\bar \theta_i, \bar k_i)$ operations and the identity operator in each $\{\bar \theta_i, \bar k_i \}$ dependent subspace.   For this, instead of defining, by integration, single party operators,  as  $\hat \Gamma_{\alpha}^{(d)}$, we can define bipartite operators in the form 
\begin{widetext}
\begin{equation}\label{bipartite}
\hat \Gamma_{\alpha_1, \alpha_2}^{d_1, d_2}=\int_0^{\frac{2\pi}{d_1}}\int_0^{\frac{2\pi}{d_2}} \int_0^{1} \int_0^{1}{\rm d}\bar \theta_1{\rm d}\bar \theta_2{\rm d}\bar k_1 {\rm d}\bar k_2 \zeta_{\alpha_1,\alpha_2}^{(d_1,d_2)}(\bar \theta_1, \bar k_1,\bar \theta_2, \bar k_2)\hat \gamma_{\alpha_1}^{(d_1)}(\bar \theta_1, \bar k_1)\otimes \hat \gamma_{\alpha_2}^{(d_2)}(\bar \theta_2, \bar k_2),
\end{equation}
\end{widetext}
where function $\zeta_{\alpha_1,\alpha_2}^{(d_1,d_2)}(\bar \theta_1, \bar k_1,\bar \theta_2, \bar k_2)$ must obey, for $i=1,2$ independently, the same periodicity conditions as functions $\zeta_{\alpha}^{(d)}(\bar \theta, \bar k)$ in Eq. (\ref{SUd}). Operator (\ref{bipartite}) can be an entangling gate, if function $\zeta_{\alpha_1,\alpha_2}^{(d_1,d_2)}(\bar \theta_1, \bar k_1,\bar \theta_2, \bar k_2)$ is not separable, and by composing such gates, one can generalize to continuous variables systems  quantum controlled logic gates \cite{example}, as well as entanglement tests and criteria that are designed for discrete variables, as the concurrence \cite{WOOTTERS}, the Peres--Horodecki criterium \cite{PERES, HORO} and the Schmidt decomposition. Of course, one can also think of generalizing  Eq. (\ref{bipartite}) to the $N>2$ partite case, leading to the possibility of more general continuous quantum state manipulations. 

 In conclusion, we have  shown how to define observables  with a continuous spectrum that can be used  to realize, over continuous variables, operations analogous to the ones defined for qubits and qudits. This is achieved through the continuous discretization of the modular variables basis. Our formulation can be applied to continuous quantum systems with either bounded or unbounded spectrum. We presented some examples of application of our results in Bell--type inequalities and quantum state measurement, comparing them to other  methods currently used to deal with continuous variables. We also show how this procedure can be extended to  multi--partite systems, and consequently be used to define, in continuous variables systems, different types of quantum protocols and algorithms. Our results open the path to a new way of dealing with quantum information and quantum state manipulation in continuous variables. 
 
\medskip

\begin{widetext}
\section*{Supplementary Information}

\section{Commutation relations and expressing states in an alternative basis} 

In order to preserve the canonical
commutation relations of $\hat{x}$ and $\hat{p}$, and consequently
of the dimensionless operators $\hat{\theta}=2\pi \hat n+\hat {\bar \theta}$ and $\hat{k}=\hat m+\hat {\bar k}$, 
given
by $[\hat{\theta},\hat{k}]=2\pi i$, the integer and modular operators
must satisfy specific commutation relations \cite{RUSSOS}. In particular, we have that  $[\hat{\bar{\theta}},\hat{\bar{k}}]=0$, implying the existence of a common basis where both operators are diagonal. Moreover, this basis is complete, and we can use it as an alternative basis to express state in momentum or position representation:  
\begin{eqnarray}
&&\ket{\theta}=\int_{0}^{1}\ud\bar{k}'e^{-\I\bar{k}'\bar{\theta}/2}e^{-\I\left(\theta-\bar{\theta}\right)\bar{k}'}\ket{\left \{\bar{\theta},\bar{k}\right \}} \nonumber \\
&&\ket{k}=\frac{1}{\sqrt{2\pi}}\int_{0}^{2\pi}\ud\bar{\theta}'e^{+\I\bar{k}\bar{\theta}'/2}e^{+\I\left(k-\bar{k}\right)\bar{\theta}'}\ket{\left \{\bar{\theta},\bar{k}\right \}}
\end{eqnarray}

A general quantum state $\int g(\theta)\ket{\theta}{\rm d}\theta$
can be written as $\sum_{n}\int_{0}^{1}\int_{0}^{2\pi}{\rm d}\bar{k}{\rm d}\bar{\theta}e^{\frac{i\bar{\theta}\bar{k}}{2}}e^{in\bar k}g(n+\bar{\theta})\ket{\left \{\bar{\theta},\bar{k}\right \}}=\int_{0}^{1}\int_{0}^{2\pi}{\rm d}\bar{k}{\rm d}\bar{\theta}\tilde g(\bar \theta, \bar k)\ket{\left \{\bar{\theta},\bar{k}\right \}}$.
The definition of the length scale ($l=1$ in the reduced coordinates $\hat \theta$ and $\hat k$) is, in principle, arbitrary. In order to simplify notation, we can assume that the
width of the distribution $g(\theta)$ is much smaller than $1$, so that only states for which $n=0$  are considered and the sum is neglected.  Notice that this
choice can be made for any distribution and that it also defines
the spacing between different values of $\hat{m}$, since they depend
on $1/l=1$ in the reduced coordinates. Thus, we can express any experimentally produced quantum state
simply as $\int g(\theta)\ket{\theta}{\rm d}\theta=\int_{0}^{1}\int_{0}^{2\pi}{\rm d}\bar{k}{\rm d}\bar{\theta}e^{\frac{i\bar{\theta}\bar{k}}{2}}g(\bar{\theta})\ket{\left \{\bar{\theta},\bar{k}\right \}}$. 

We stress however that making this assumption is equivalent to considering only a subspace of the Hilbert space. In practice, experiments can only access subspaces of the Hilbert space of infinite dimensions. Thus, states can always be expressed as the ones spanned by the $n=0$ subspace. 

\smallskip
\section{ Some useful quantum operations} 

Phase-shift gates are defined
by \cite{BVL}:

\begin{align*}
\hat X(\Phi)\ket\theta=\ket{\theta+\Phi}\qquad & \hat X(\Theta)=\exp\left(\I\Theta\hat{k}\right)\\
\hat Z(K)\ket k=\ket{k+K}\qquad & \hat Z(K)=\exp\left(\I K\hat{\theta}\right)
\end{align*}

As shown in \cite{TASCA},  the above operations can be implemented using standard optics elements.

An important tool to implement operators as $\hat S_{\alpha}^{(d)}(\bar \theta, \bar k)$ are quadratic gates. In optics, the free propagation is a very simple example of a quadratic gate, since at a distance $s=ct$ it is represented by the operator $e^{-\I\hat{k}^{2}\varphi}$, with $\varphi=\frac{\lambda s}{4\pi l^{2}}$,
where $c$ is the speed of light and $\lambda$ the wavelength. Using this, we can defined the gate $\hat U[\varphi]$:
\[
\hat U\left[\varphi\right]=e^{-\I\hat{k}^{2}\varphi}. 
\]
By combining  $\hat Z(K)$ and $\hat U(\varphi)$ gates, we obtain the 
transformations :

\begin{eqnarray}\label{transform}
&&\hat U\left[\frac{\varphi}{2}\right]\hat Z(1)\hat U^{\dagger}\left[\frac{\varphi}{2}\right]\ket{\left\{ \bar{\theta},\bar{k}\right\} }= \\
&&e^{-\I\bar{k}\bar{\theta}/2}e^{-\I(2\bar{k}+1)\varphi}e^{\I\frac{1}{2}\bar{k}\left(\overline{\bar{\theta}+\varphi}\right)}e^{\I\left(\overline{\bar{\theta}+\varphi}\right)}\ket{\left\{ \overline{\bar{\theta}+\varphi},\bar{k}\right\} }\nonumber
\end{eqnarray}

In order to engineer well adapted  $\hat{S}_{\alpha}^{(d)}(\bar \theta, \bar k)$ operators, we first set $\varphi=2\pi/d$, that leads to $SU(d)$ non--diagonal operators from diagonal ones (created by setting $\varphi=0$). By acting on such non--diagonal operators with $\hat X(\Theta)$ and $\hat Z(K)$ with judiciously choses parameters $\Theta$ and $K$, we can create different, non--commuting, non--diagonal operators in $SU(d)$. Using the Fourier transform and exchanging the roles of $\hat X(\Theta)$ and $\hat Z(K)$ lead to analogous transformations on variable $\bar k$.

\section{General recipe for finding diagonal $\hat \Gamma_{\alpha}^{(d)}$ operators: conditions on the $F(\bar \theta, \bar k)$ function}

The $\{\bar \theta,\bar k\}$ space, for a fixed value of $\bar k$, can be mapped into a circle, as shown Fig. 1 of the main text. This circle can be divided
into $d$ domains of size $\frac{2\pi}{d}$. The basis states of each $\{\bar \theta, \bar k \}$ dependent subspace differ by displacements in variable $\bar \theta$ that are multiples of  $\frac{2\pi}{d}$, as illustrated in Fig. 1 of the main text.
Therefore, for a given pair $\{\bar \theta, \bar k\}$, we define the vectors $\left\{ \ket{\{\bar \theta+\frac{2\pi n}{d},\bar k\}}/n\in\left[0,d-1\right]\right\} $
as a basis of a $d$-dimensional orthonormal vector space. A diagonal
operator in this space can be written as:

\[
\hat{\Omega}(\bar \theta, \bar k)=\underset{{\scriptstyle n=0}}{\sum^{d-1}}{\cal F}_{n}(\bar \theta, \bar k)\ket{\{\bar \theta+\frac{2\pi n}{d}, \bar k\}}\bra{\{\bar \theta+\frac{2\pi n}{d}, \bar k \}},
\]
where $\bar \theta \in [0,2\pi/d [$.   At the same time, an operator should be defined on the whole circle, \textit{i.e.}
:

\begin{equation}\label{delta}
\hat{\Delta}= \int_0^{2\pi} d \bar \theta d \bar k F(\bar \theta, \bar k) \ket{\{\bar \theta, \bar k \}}\bra{\{\bar \theta, \bar k \}}=  \int_0^{2\pi/d}d \bar \theta d\bar k \hat{\Omega}(\bar \theta, \bar k)= \intop_{0}^{2\pi/d} d\bar \theta d\bar k \underset{{\scriptstyle n=0}}{\sum^{d-1}}c_n{\cal F}(\bar \theta, \bar k)\ket{\{\bar \theta+\frac{2\pi n}{d}, \bar k\}}\bra{\{\bar \theta+\frac{2\pi n}{d}, \bar k \}},
\end{equation}
%F\text{(\ensuremath{\bar \theta})}\ket{\bar \theta}\bra{\bar \theta},
where $c_{n}$ is a complex coefficient. The first term on the right of the above equation can always be written as follows: 
\begin{equation}\label{F}
\int_0^{2\pi} d \bar \theta d \bar k F(\bar \theta, \bar k) \ket{\{\bar \theta, \bar k \}}\bra{\{\bar \theta, \bar k \}}=\intop_{0}^{\frac{2\pi}{d}} d\bar \theta d\bar k\underset{{\scriptstyle n=0}}{\sum^{d-1}}F\left(\ensuremath{\bar \theta}+\frac{2\pi n}{d}, \bar k\right)\ket{\{\bar \theta+\frac{2\pi n}{d},\bar k \}}\bra{\{\bar \theta+\frac{2\pi n}{d}, \bar k\}}
\end{equation}

A diagonal $\hat \Gamma_{\alpha}^{(d)}$ operator is such that (\ref{delta})=(\ref{F}). For this,  function $F$ needs to have certain periodicity properties, so that $F\left(\ensuremath{\bar \theta}+\frac{2\pi n}{d}, \bar k \right)={\mathcal F}_n\left(\ensuremath{\bar \theta}, \bar k\right)=c_{n}\mathcal{F}\left(\ensuremath{\bar \theta}, \bar k\right)$. We see that for a fixed $\bar k$, this condition only applies to the $\bar \theta$ dependency of the function. Thus, in the following discussion, we will neglect the $\bar k$ variable and include it after reaching the main conclusions. In order to satisfy the conditions above, we have that $F\left(\ensuremath{\bar \theta}+\frac{2\pi n}{d} \right)=c_{n}\mathcal{F}\left(\ensuremath{\bar \theta}\right)$ must be such that:

\[
F\left(\ensuremath{\bar \theta} \right)=\acco{\begin{aligned}c_{0}\mathcal{F}(\bar \theta) & \ \text{if }\ \bar \theta\in\left[0,\frac{2\pi}{d}\right]\\
\vdots\\
c_{n}\mathcal{F}(\bar \theta) & \ \text{if }\ \bar \theta\in\left[\frac{2\pi n}{d},\frac{2\pi(n+1)}{d}\right]\\
\vdots
\end{aligned}
}
\]

Also, as the splitting of the physical operator onto the $d$ domains of size $2\pi/d$ should not introduce irregularities or discontinuities in the function $F(\bar \theta)$, it
has to be at least continuous at the edges of each domain, resulting
in the following conditions on the $\left\{ c_{n}\right\} $ coefficients:

\begin{equation}
\acco{\begin{aligned}c_{0}\mathcal{F}\left(\frac{2\pi}{d} \right) & =c_{1}\mathcal{F}\left(0\right)\\
c_{1}\mathcal{F}\left(\frac{2\pi}{d} \right) & =c_{2}\mathcal{F}\left(0 \right)\\
\vdots\\
c_{d-1}\mathcal{F}\left(\frac{2\pi}{d} \right) & =c_{0}\mathcal{F}\left(0\right)
\end{aligned}
}
\end{equation}

From those, we deduce that the $c_{n}$'s are the $d$th roots of
unity. Moreover, all the $\frac{c_{n+1}}{c_{n}}$ quotients
have to be equal to the same $d$--th root of unity:

\begin{equation}\label{S1}
\frac{c_{n+1}}{c_{n}}=e^{\I\frac{2\pi }{d}}
\end{equation}

It is clear that Eq. (\ref{S1}) is simply one possible solution, and that  for each $n$ , there are $d$ solutions $c_{n}^{(l)}$,
where the label $l$ belongs to $\left\{ 0,1\cdots,d-1\right\} $.  However, in this section, for the sake of clarity, we will only consider one solution. 

By convention, we choose $c_{0}=1$, corresponding only to a global
phase and we obtain: 
\begin{equation}
c_{n}=e^{\frac{2\pi in}{d}}
\end{equation}
\begin{equation}
\begin{aligned}F(\bar \theta+\frac{2\pi n}{d})=e^{\frac{2\pi i n}{d}} & {\cal F}(\bar \theta)\quad;\forall \ \bar \theta\in[0,2\pi]\end{aligned}
\end{equation}

From the above conditions, we determine the properties a diagonal
operator should have in order to admit a decomposition as a continuum of $SU(d)$--type operators in the whole circle.

Taking as an example $d=2$, with $c_{0}=1$, we have that  
$c_{1}=-1$. One of the simplest, most regular $\mathcal{F}$ function
verifying :

\begin{equation}
\forall\ \bar \theta\in\left[0,2\pi\right],F\left(\ensuremath{\bar \theta}+\pi\right)=-F\left(\ensuremath{\bar \theta}\right)
\end{equation}
is the cosine function, so $F(\bar \theta)=\cos(\bar \theta)$. 

Since the above discussion simply does not depend on  $\bar k$, this variable can de freely reincorporated to the functions $F(\bar \theta)$  and ${\cal F}(\bar \theta)$ simply by retransforming them into $F(\bar \theta, \bar k)$ and  ${\cal F}(\bar \theta, \bar k)$. We see that, under these conditions,  $F(\bar \theta, \bar k)\equiv\zeta_{\alpha}^{(d)}(\bar \theta, \bar k)$ for $\alpha$ such that $\hat \Gamma_{\alpha}^{(d)}$ is diagonal. In particular,  as shown in the detailed discussion on the $SU(2)$--type operators made in Section III and the discussion on possible experimental realizations of the proposed operators made in Section II, we see that functions $\zeta_{\alpha}^{(d)}(\bar \theta, \bar k)=\cos{(\bar \theta -\bar k \pi)}$ are the most adapted solutions in the $SU(2)$--type case. These solutions clearly satisfies the conditions imposed above as well. It is important to notice that the conditions state above are very loose, providing a large freedom of choice for $F(\bar \theta, \bar k)$, according to the envisaged application. 

From the diagonal operators one can use displacements in the modular basis and define non--diagonal ones. The main goal is to be able to define, for a given $d$, $d^2-1$ independent matrices in each $\{ \bar \theta, \bar k \}$ dependent subspace. It is clear that there is a large freedom of choice on the type of operation one uses to reach this goal, even when imposing that all $\hat \Gamma_{\alpha}^{(d)}$ must be traceless and hermitian. In the present work, we will focus on operations that can be realized with current technology in experimental set--ups. We will thus focus on a specific experimental set--up that has already been used in entanglement tests using modular variables \cite{RUSSOS, STEVE}. It consists of the transverse coordinates of single photons. This choice restricts the space of allowed operations, as is the case for any experimental set--up and its natural constraints \cite{TASCA, BLV}. However, we show in the next section that it is possible to create  $\hat \Gamma_{\alpha}^{(d)}$ operators of arbitrary dimensions using the available operations in this type of set--up. 

\section{Realizing modular operators experimentally using the transverse degrees of freedom of photons}

An important aspect of our results is the fact that they are well adapted to an immediate experimental implementation by manipulating the transverse spatial degrees of freedom of photons using available linear optical elements. We present several methods to implement the modular operators in this context.

We first consider measurements of arbitrary operators of the type $\hat{\Gamma}_{\alpha}^{(d)}$, as described in the main text. The basic idea is shown in Fig. \ref{fig2}.  Using an interferometer, analogous to the one introduced in \cite{SAULO}, we observe the detection probability at the two outputs $1$ and $2$.  We assume that the ``beam splitters" are 50/50, and antisymmetric.  As previously introduced,  $\hat D_{\alpha}^{(d)}$ are diagonal operators and  $\hat S_{\alpha}^{(d)}$ are displacement operators such that  $\hat \Gamma_{\alpha}^{(d)}=2{\rm Re}\left [ \hat S_{\alpha}^{(d)}\hat D_{\alpha}^{(d)}\right ]$ are observables with a continuum spectrum consisting of a infinite sum of $SU(d)$ operators acting on $d$ regions of $\bar \theta$, as in Fig. 1 of the main text. 

We suppose that coordinates $\hat x$ and $\hat p$, or equivalently,  $\hat \theta$ and $\hat k$ refer to the transverse position and momentum of a single photon. Thus, a general quantum state of the transverse momentum (or position) of the photon can be written in the $\bar \theta$ and $\bar k$ basis, as shown in Section I. In this scenario, operators  $\hat D_{\alpha}^{(d)}$  and $\hat S_{\alpha}^{(d)}$ can be engineered by combining Spatial Light Modulators (SLMs), lenses (that can perform complete, or fractional, Fourier transforms), and free propagation, for instance.  A detailed description of the implementation of continuous variable quantum logic gates using these linear optical elements has been provided \cite{TASCA}.

\begin{figure}
 \centerline{\includegraphics[width=.6 \columnwidth]{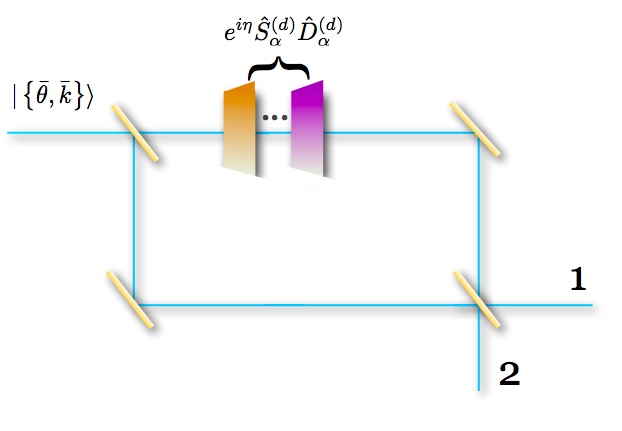}
 }
\caption{ Principle of the Mach--Zender type interferometer  leading to the experimental production of a continuum of $SU(d)$--type operators acting on the transverse coordinates of a single photon. The operation $e^{i\eta}\hat S_{\alpha}^{(d)}\hat D_{\alpha}^{(d)}$ is realized in one of the arms of the interferometer by combining different linear optics elements and the free propagation. The phase $\eta$ can be chosen so one can access either the real or imaginary parts of $\langle \hat S_{\alpha}^{(d)}\hat D_{\alpha}^{(d)} \rangle$ when measuring the difference between the detection probabilities in each port, $P_1-P_2$. 
 \label{fig2}}
\end{figure}

In this scenario, the phase $\eta$ between the two arms can be controlled by a dephasing element.  Choosing $\eta=0$, we have that the output probabilities are given by: 
\begin{equation}
P_j = \frac{1}{2} \left(1 +(-1)^j  \langle \hat S_{\alpha}^{(d)} \hat D_{\alpha}^{(d)} + \hat D_{\alpha}^{(d) \dagger} \hat S_{\alpha}^{(d)\dagger} \rangle\right)
\end{equation}
where $j=1,2$. By taking the difference between photon counts in each exiting port of the interferometer, we have that 
\begin{equation}
P_1 - P_2 = \langle \mathrm{Re} \left [ \hat S_{\alpha}^{(d)} \hat D_{\alpha}^{(d)}\right ] \rangle
\end{equation}
Choosing rather $\eta = \pi/2$, we have
\begin{equation}
P_1 - P_2 = \langle \mathrm{Im}\left [\hat S_{\alpha}^{(d)} \hat D_{\alpha}^{(d)}\right] \rangle
\end{equation}
\par

Thus, it is clear that by choosing  $\hat S_{\alpha}^{(2)} \hat D_{\alpha}^{(2)}$ as in Section III, which is a realistic choice  based on operations currently realized to manipulate the transverse coordinates of single photons \cite{TASCA}, we can experimentally obtain $\langle \hat \Gamma_{\alpha}^{(2)} \rangle=P_1-P_2$ simply by measuring the difference between photon counts in each output port of the interferometer depicted in Fig. {\ref{fig2}}.  This experimental configuration is well adapted to protocols such as quantum state or process tomography, where expected values of the $\hat \Gamma_{\alpha}^{(2)}$ operators are measured. Also, in Bell--type inequalities, one needs to compute correlations between operators  $\hat \Gamma_{\alpha}^{(2)}$ acting in two different subsystems. This can be done by measuring the coincidence counts in two interferometers as the one depicted in Fig. {\ref{fig2}}. 

In the case where one want to obtain states resulting from the action of operators $\hat \Gamma_{\alpha}^{(2)}$, as for instance the equivalent of continuous $SU(2)$ rotations or, in the many party case, conditional quantum logical operations or quantum algorithms, the architecture presented in Fig. \ref{fig2a} can be used. In this case, linear optical elements should be added to the lower arm of the interferometer as well. Such operations are  are the perfect conjugate to the ones implemented in the upper arm, so as in each exit, one obtain different  $\hat \Gamma_{\alpha}^{(2)}$ operators, according to the choice of phase $\epsilon$. 

\begin{figure}
 \centerline{\includegraphics[width=.8 \columnwidth]{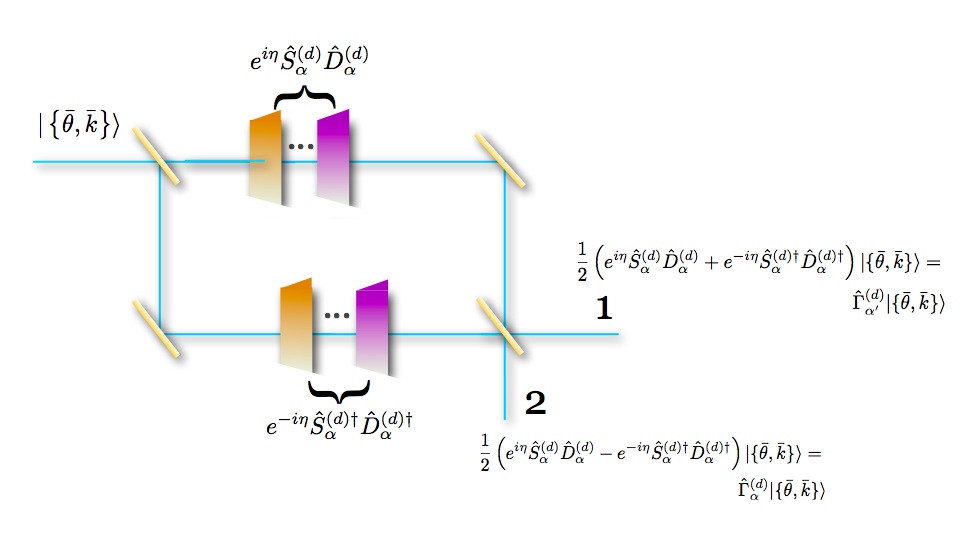}
 }
\caption{  Mach--Zender type interferometer  leading to the experimental production of a continuum of $SU(d)$--type operators acting on the transverse coordinates of a single photon. The operation $e^{i\eta}\hat S_{\alpha}^{(d)}\hat D_{\alpha}^{(d)}$ is realized in one of the arms of the interferometer by combining different linear optics elements and the free propagation.  In another arm, the conjugate operation, $\hat D_{\alpha}^{(d)\dagger}\hat S_{\alpha}^{(d)\dagger}$ is realized, up to a phase. The phase $\eta$ can be chosen so that in  each port we have states $ \hat S_{\alpha}^{(d)}\hat D_{\alpha}^{(d)} \pm \hat D_{\alpha}^{(d)\dagger}  \hat S_{\alpha}^{(d)\dagger}\ket{\{\hat \theta, \hat k \}}$ in the exit ports, up to a normalization constant.
 \label{fig2a}}
\end{figure}

From these architectures, it is clear that one can choose different reflection and transmission coefficients of the beam splitters, as well as different optical elements in each arm of the interferometer, so that instead of producing operators in the form $\hat \Gamma_{\alpha}^{(2)}$, one can produce arbitrary linear combinations of these operators, such as $\cos{\beta}\hat \Gamma_{\alpha}^{(2)}+e^{i\epsilon}\sin{\beta}\hat \Gamma_{\alpha'}^{(2)}$, with $\alpha \neq \alpha'$. 

We have seen in Section II that $\hat \Gamma_{\alpha}^{(2)}$ and $\hat \Gamma_{\alpha}^{(3)}$ operators can be obtained using the $\hat X(\Phi)$, $\hat Z(\Theta)$ and $\hat U(\phi)$ gates, that are all univocally related to specific linear optical elements:   $\hat X(\Phi)$ is a linear dephasing with slope $\Phi$ in the $\hat k$ space, that can be seen as a displacement in $\hat \theta$ space.   Equivalently, $\hat Z(\Theta)$ is a linear dephasing with slope $\Theta$ in the $\hat \theta$ space, that can be seen as a displacement in $\hat k$ space. Both can be produced with a judiciously programed Spatial Light Modulator (SLM) combined to Fourier gates, consisting of linear optical elements, as lenses and the free propagation, or other SLMs, that perform the Fourier transform of a given transverse distribution. Finally,  $\hat U(\phi)$ is the free propagation, a quadratic gate that can also be implemented/compensated, with the help of SLMs. 

The interferometric scheme presented can also be generalized so as to create operators $\hat \Gamma_{\alpha}^{(3)}$, since as shown in Section IV, such operators can be realized by a combination of $\hat X(\Phi)$, $\hat Z(\Theta)$ and $\hat U(\phi)$ gates. In general, such operations create displacements such that it is always possible to create a set of $d^2-1$ independent matrices leading to the definition of $SU(d)$--type operators in $\bar \theta \bar k$ dependent subspaces. 

In the next subsection, we will present a systematic method of creating arbitrary operators $\hat \Gamma_{\alpha}^{(d)}$ that has the advantage of being geometrically intuitive. It is based in the same interferometric architecture as the one presented in  Fig. \ref{fig2}. However, operators $\hat S_{\alpha}^{(d)}\hat D_{\alpha}^{(d)}$ are created by multiple reflexions of the transverse distribution with respect to well chosen points in the transverse plane.

\subsubsection{$SU(d)$ generators by reflections}
%%%%%%%%%%%%%%%%%%%%%%%%%%%%%%%%%%%%%%%%%%%%%%%%%%
\begin{figure}
\includegraphics[width=0.8\textwidth]{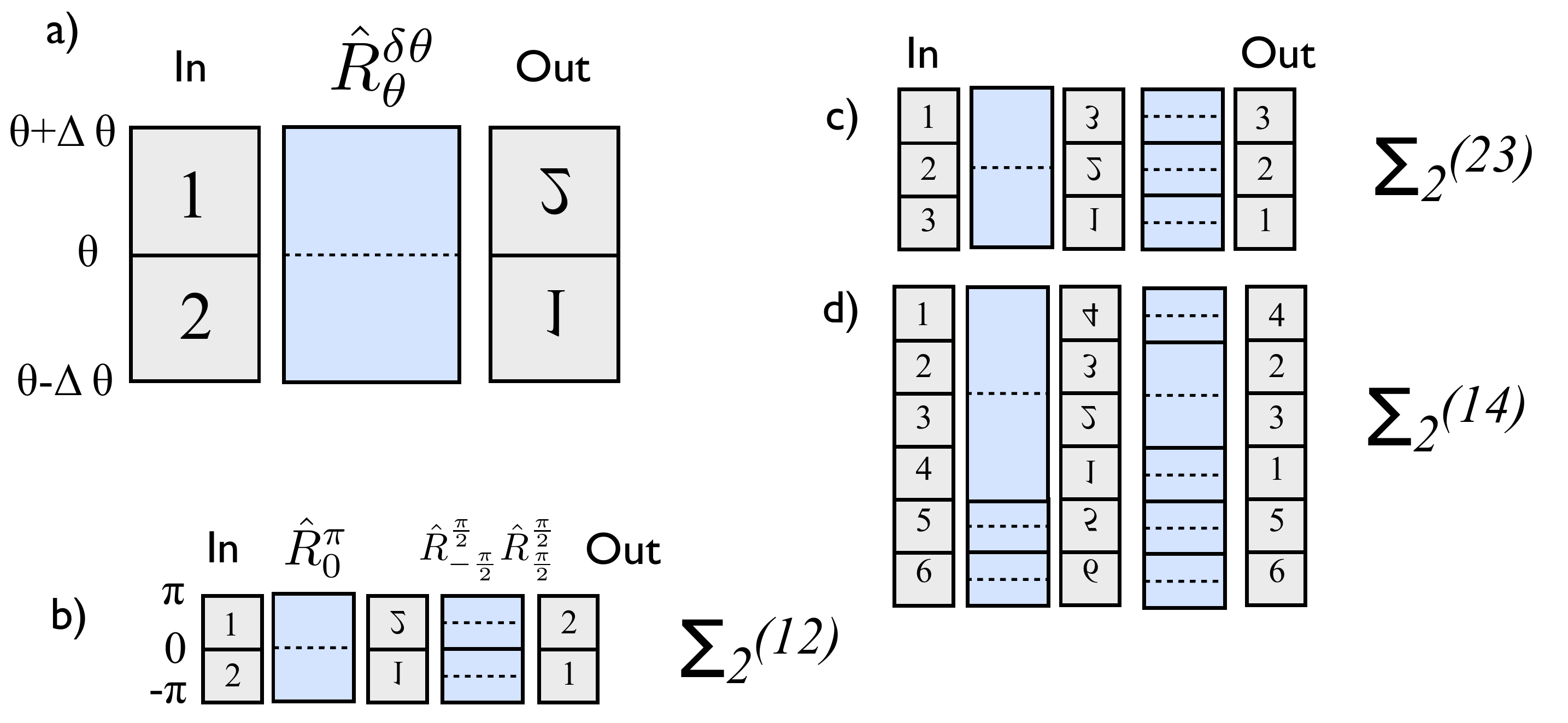}
\caption{  (a) Reflection operator (blue box) reflects the region $2 \delta\theta$ around the axis (dotted line) located at $\theta_p$. (b) Three reflection operators can implement the operator $\hat {\Sigma}_2^{(12)}$. (c) Example of combinations of reflections leading to $\hat \Sigma_{2}^{13}$ for $d=3$, (d) Example of combinations of reflections leading to $\hat \Sigma_{2}^{14}$ for $d=6$.}
\label{fig:reflect}
\end{figure}
%%%%%%%%%%%%%%%%%%%%%%%%%%%%%%%%%%%%%%%%%%%%%%%%%%
\par
All $SU(d)$ generators can be expressed as being  either diagonal or equivalent to a Pauli operator acting on a two-dimensional subspace of the overall $d$ dimensional subspace. We have seen that the diagonal operators can be implemented using an SLM, which can be programmed to apply a position-dependent phase to different regions of an optical field.  To construct the non-diagonal operators, we will make use of the decomposition in terms of $2 \times 2$ Pauli matrices acting on subspaces of the $d$-dimensional system.
\par
We show now is that these operators can be constructed in the  $\{\bar \theta, \bar k \}$ basis by manipulating the $ \theta$ variables using a series of reflection operators.  In the next subsection we show how the reflection operators can be implemented using optics. Let us define the reflection operator $\hat {R}_{\theta_p}^{\delta \theta}$.  The point $\theta_p$ defines the symmetry axis of the reflection, and $\delta \theta$ defines the overall region in which the operator acts, as illustrated in Fig. \ref{fig:reflect} a).  The operator $\hat {R}_{\theta_p}^{\delta \theta}$ can be written as
\begin{equation}
\hat {R}_{\theta_p}^{\delta \theta} = \int\limits_{\theta_p-\delta\theta}^{\theta_p+\delta\theta} \ket{2 \theta_p - \theta^\prime}\bra{\theta^\prime} d\theta^\prime, 
\end{equation} 
which clearly takes $\theta^\prime$ into $2 \theta_p-\theta^\prime$ when $\theta^\prime \in [ \theta_p-\delta\theta, \theta_p-\delta\theta ]$. 
\par
Any generalized Pauli operator can be written as a series of operators $\hat {R}_{\theta_p}^{\delta \theta}$ acting on different subspaces.  In order to illustrate this, let us define the notation $\hat {\Sigma}_i^{(jk)}$, where $j,k$ refer to the two particular regions which will be acted upon by  two-dimensional Pauli operators $\hat {\sigma}_i(\theta)$ such that: 
\begin{eqnarray}\label{pauliref}
&&\hat  {\sigma}_1(\theta)=\ket{\theta}\bra{\theta}-\ket{\theta^\prime}\bra{\theta^\prime} \nonumber \\
&&\hat  {\sigma}_2(\theta)=\ket{\theta}\bra{\theta^\prime}+\ket{\theta^\prime}\bra{\theta} \\
&&\hat  {\sigma}_3(\theta)=i\ket{\theta}\bra{\theta^\prime}-i\ket{\theta^\prime}\bra{\theta} \nonumber 
\end{eqnarray}  
where $\theta$ is defined in the $j$ region and $\theta^\prime$ in the $k$ region.

We start by considering $\int_{\theta_p-\delta \theta}^{\theta_p+\delta \theta}\hat {\sigma}_2(\theta)d\theta$, that can be identified to operator $\hat {\Sigma}_2^{(12)}$, as illustrated in Fig. \ref{fig:reflect} b).  We can see that    
\begin{equation}
\int_{\theta_p-\delta \theta}^{\theta_p+\delta \theta}\hat {\sigma}_2(\theta)d\theta = \hat {\Sigma}_2^{(12)} = \hat {R}_{-\frac{\pi}{2}}^{\pi} \hat {R}_{\frac{\pi}{2}}^{\pi}\hat {R}_{0}^{2 \pi} 
\end{equation}
Thus, in order to create operator $\hat \Gamma_{2}^{(2)}$, we can identify $\hat S_{2}^{(2)}\equiv \hat {\Sigma}_2^{(12)}$ and $\hat D_{\alpha}^{(d)} \equiv e^{i\hat \theta}$. Thus, in the output ports of the interferometer depicted in Fig. \ref{fig2}, we have either $\langle {\rm Re} \left [\hat \Gamma_{2}^{(2)}\right ] \rangle$ or  $\langle {\rm Im} \left [\hat \Gamma_{2}^{(2)}\right ] \rangle$, as in the previous section. Equivalently, by using the interferometer depicted in Fig. \ref{fig2a}, we can obtain $\hat \Gamma_{2}^{(2)}\ket{\psi}$.  Thus, we can obtain $\hat \Gamma_{2}^{(2)}\ket{\psi}$ by sending the state $\ket{\psi}$ through the reflection scheme depicted in Fig. \ref{fig:reflect} b). To measure the expection value of this operator, we can combine the reflection scheme with the interferometer shown in Fig. \ref{fig2}.  At the outputs we have either $\langle {\rm Re} \left [\hat \Gamma_{2}^{(2)}\right ] \rangle$ or  $\langle {\rm Im} \left [\hat \Gamma_{2}^{(2)}\right ] \rangle$, as in the previous section. 

Implementing $\hat \Gamma_{3}^{(2)}$ requires only the application of an additional phase on the $j$ and $k$ regions, which can be done with phase elements inserted between reflections.  Theses phase elements are diagonal in the $\theta$ basis, and can be implemented using an SLM, for example.

Similar examples for $d=3$ and $d=6$ are shown in Fig. \ref{fig:reflect} (c) and (d). They show that it is indeed possible to create, through reflections, the equivalent of  an infinite sum of Pauli operators acting on a two-dimensional subspace of the overall $d$ dimensional subspace, i.e., the $\hat {\Sigma}_i^{(jk)}$ operators. These operators, combined to diagonal ones with properties satisfying the conditions detailed in Section I, lead to arbitrary $\hat \Gamma_{\alpha}^{(d)}$ operators in a systematic way.

\subsubsection{Optical implementation of reflections}

\par

%%%%%%%%%%%%%%%%%%%%%%%%%%%%%%%%%%%%%%%%%%%%%%%%%%
\begin{figure}
\includegraphics[width=0.6\textwidth]{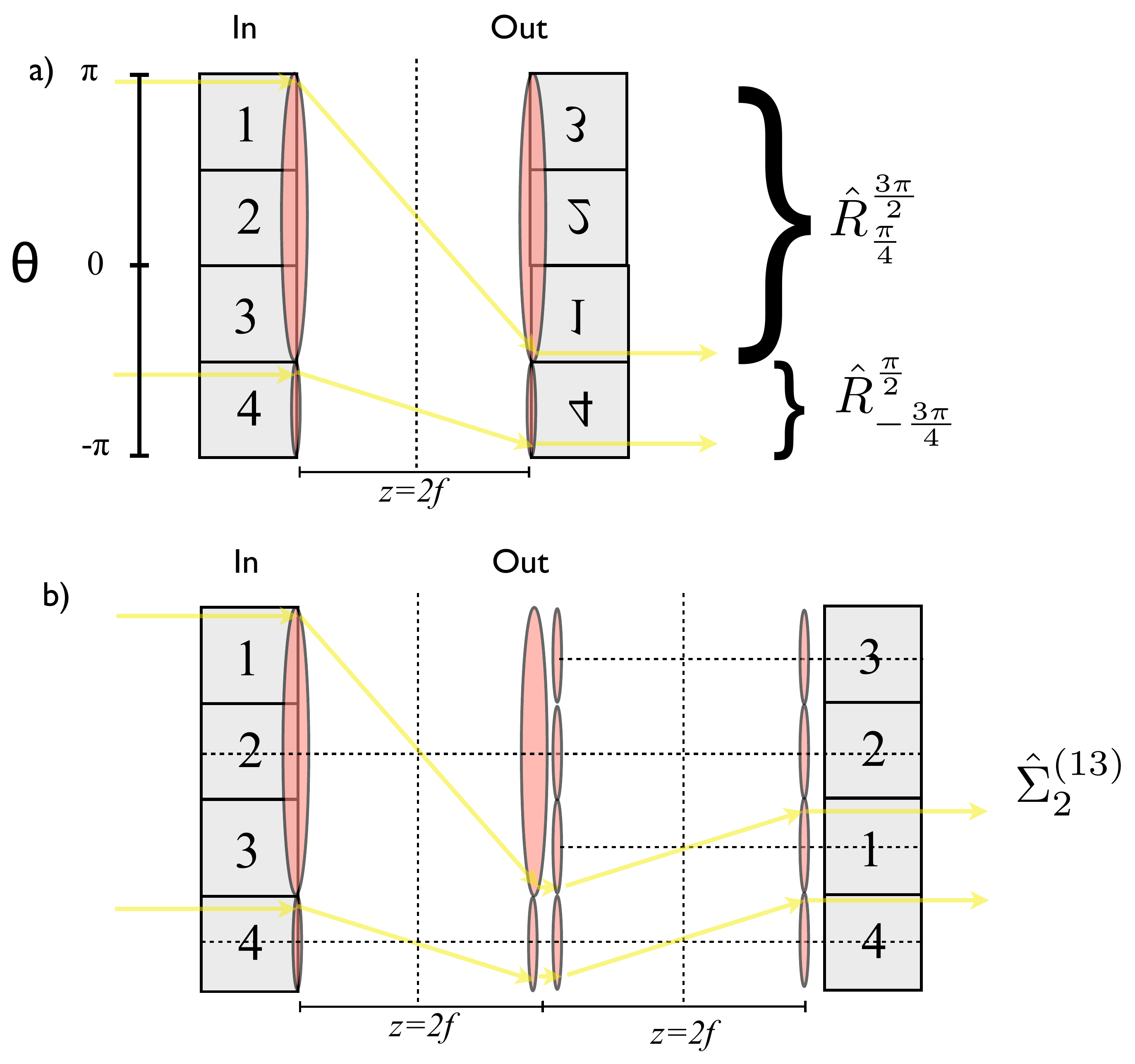}
\caption{a) Lens systems to implement the $\oper{R}_{\frac{\pi}{4}}^{\frac{3\pi}{2}}$ and $\oper{R}_{\frac{-3\pi}{4}}^{\frac{\pi}{2}}$ reflection operators. b) Using quadratic phases to implement the operator $\oper{\Sigma}_2^{(13)}$ for $d=4$.}
\label{fig:system}
\end{figure}
%%%%%%%%%%%%%%%%%%%%%%%%%%%%%%%%%%%%%%%%%%%%%%%%%%
The implementation of these reflection operators is simple in may cases.  For example, consider the transverse spatial degree of freedom $x$ of a photon, and define dimensionless $\theta$ accordingly. Then, $\hat {R}_{\theta_p}^{\delta \theta}$ is simply an optical imaging system centered at $\theta_p$ that acts only in the region $\delta \theta$.  It is known that a $4f$ imaging system produces an inverted image of the object, which is exactly the necessary reflection operation around the axis $\theta_p$. This can be implemented with the lens systems shown in Fig. \ref{fig:system} (a), showing how to produce  $\hat{R}_{\frac{\pi}{4}}^{\frac{3\pi}{2}}$ and $\hat {R}_{-\frac{3\pi}{4}}^{\frac{\pi}{2}}$.  In red, we show quadratic phase elements such as lenses. More realistically such quadratic phases are more easily produced by using SLMs, which can be programmed to apply a quadratic phase only to certain regions of $\theta$.  The quadratic phase corresponds to a lens with focal length $f$, and the distance between the lenses is $z=2f$.  The free propagation in also represented by a quadratic phase of the form $\exp(i f \hat{p}^2/|p|)$ \cite{TASCA}. Using quadratic phases to implement the operator $\hat {\Sigma}_2^{(13)}$ is shown in Fig. \ref{fig:system} (b) for $d=4$.      

\section{Detailed description and interpretation of the $SU(2)$--type operators}

The $\bar \theta$, $\bar k$ dependent basis is formed by pairs of  states $\ket{\left\{ \bar{\theta},\bar{k}\right\} }$ and $\ket{\left\{ \bar{\theta}-\pi,\bar{k}\right\} }$ in the circle (see Fig. 1 (a) in the main text). By considering two diametrically opposed states $\ket{\theta}$ and $\ket{\theta-\pi}$, we have that, in the modular basis, they relate as: 
\begin{equation}
\acco{\begin{aligned}\ket{\theta} & =\int_{0}^{1}d\bar{k}e^{-\I\bar{k}\bar{\theta}/2}\ket{\left\{ \bar{\theta},\bar{k}\right\} }\\
\ket{\theta-\pi} & =\int_{0}^{1}d\bar{k}e^{-\I\bar{k}(\bar{\theta}-\pi)/2}\ket{\left\{ \bar{\theta}-\pi,\bar{k}\right\} }
\end{aligned}
}\text{where }\theta\in\left]0,\pi\right[
\end{equation}
(Notice that  $\bar{\theta}\in\left]0,\pi\right[\text{, }\overline{\theta-\pi}=\bar{\theta}-\pi$)

Hence, the non diagonal operators between the two modular states are:

\begin{equation}
\left\{ e^{-\I\bar{k}\frac{\pi}{2}}\ket{\left\{ \bar{\theta}-\pi,\bar{k}\right\} }\bra{\left\{ \bar{\theta},\bar{k}\right\} },e^{\I\bar{k}\frac{\pi}{2}}\ket{\left\{ \bar{\theta},\bar{k}\right\} }\bra{\left\{ \bar{\theta}-\pi,\bar{k}\right\} }\right\} 
\end{equation}
Using the operations defined in Section II of Supplementary Material and the interferometric set--up introduced in Section II, we can compute 
$\hat U\left[\varphi\right]Z(K)U^{\dagger}\left[\varphi\right]$ with
$K=\pm1$ and $\varphi=\frac{\pi}{2}$. We obtain that:

\begin{align}\label{pauli}
\hat U\left[\frac{\pi}{2}\right]\left(\hat Z(1)+\hat Z(-1)\right)\hat U^{\dagger}\left[\frac{\pi}{2}\right] & =\int_{0}^{1} d\bar{k}\intop_{0}^{\pi} d\bar{\theta}e^{\I\bar{\theta}}e^{-\I\bar{k}\pi}\left[\I e^{-\I\bar{k}\frac{\pi}{2}}\ket{\left\{ \bar{\theta}-\pi,\bar{k}\right\} }\bra{\left\{ \bar{\theta},\bar{k}\right\} }+h.c.\right] \nonumber \\
 & \phantom{=}+e^{-\I\bar{\theta}}e^{\I\bar{k}\pi}\left[\I e^{-\I\bar{k}\frac{\pi}{2}}\ket{\left\{ \bar{\theta}-\pi,\bar{k}\right\} }\bra{\left\{ \bar{\theta},\bar{k}\right\} }+h.c.\right]\\
 & =2\int_{0}^{1}d\bar{k}\intop_{0}^{\pi}d\bar{\theta}\cos\left(\bar{\theta}-\bar{k}\pi\right)\left[\I e^{-\I\bar{k}\frac{\pi}{2}}\ket{\left\{ \bar{\theta}-\pi,\bar{k}\right\} }\bra{\left\{ \bar{\theta},\bar{k}\right\} }+h.c.\right] \nonumber
\end{align}

Writing (\ref{pauli}) in its matrix form in the basis 
\begin{equation}\label{basis}
\left\{ e^{-\I\bar{k}\frac{\pi}{2}}\ket{\left\{ \bar{\theta}-\pi,\bar{k}\right\} },\ket{\left\{ \bar{\theta},\bar{k}\right\} }\right\},
\end{equation}
we have: 

\begin{eqnarray}\label{sigmay}
\hat U\left[\frac{\pi}{2}\right]\left(\hat Z(1)+\hat Z(-1)\right)\hat U^{\dagger}\left[\frac{\pi}{2}\right] & =2\int_{0}^{1}d\bar{k}\intop_{0}^{\pi}d\bar{\theta}\cos\left(\bar{\theta}-\bar{k}\pi\right)\begin{pmatrix}0 & \I\\
-\I & 0
\end{pmatrix}(\bar \theta, \bar k)
\end{eqnarray}
that is the integral of Pauli operators over the whole circle, since the matrix in Eq.  (\ref{sigmay}) can be written as $\hat \sigma_3(\bar \theta, \bar k)$, i.e., a $\bar \theta, \bar k$ dependent Pauli matrix. 

Analogously, the other $2$ $\{\bar \theta, \bar k\}$ dependent Pauli matrices can be defined, up to a global phase in the basis state (see (\ref{basis})), as in Eq. (3) of the main text:
 \begin{eqnarray}\label{pauli2} 
&&\hat \sigma_1(\bar \theta, \bar k) \equiv \ket{\bar \theta, \bar k}\bra{\bar \theta, \bar k}-\ket{\bar \theta+\pi, \bar k}\bra{\bar \theta+\pi, \bar k} \nonumber \\
&&\hat \sigma_2(\bar \theta, \bar k) \equiv \ket{\bar \theta, \bar k}\bra{\bar \theta+\pi, \bar k}+\ket{\bar \theta+\pi}\bra{\bar \theta, \bar k} \\
 &&\hat \sigma_3(\bar \theta, \bar k) \equiv i\left (\ket{\bar \theta, \bar k}\bra{\bar \theta+\pi, \bar k}-\ket{\bar \theta+\pi}\bra{\bar \theta, \bar k} \right )\nonumber 
 \end{eqnarray}
Notice that the redefinition of the basis state does not change qualitatively our main results. However, it is at the origin of the appearance of a $\bar k$ dependence of function $F{\bar \theta, \bar k}$ defined in Section I. 

From the previous discussions, we can identify

\begin{equation}
\begin{aligned}\hat \Gamma_{1}^{(2)} & =\hat Z(1)+\hat Z(-1)=\int_{0}^{1}d\bar{k}\intop_{0}^{\pi}d\bar{\theta}\cos\left(\bar{\theta}-\bar{k}\pi\right)\hat \sigma_{1}(\bar{\theta},\bar{k})\\
\hat \Gamma_{3}^{(2)} &=  \hat U\left[\frac{\pi}{2}\right]\left(\hat Z(1)+\hat Z(-1)\right)\hat U^{\dagger}\left[\frac{\pi}{2}\right]=-\int_{0}^{1}d\bar{k}\intop_{0}^{\pi}d\bar{\theta}\cos\left(\bar{\theta}-\bar{k}\pi\right)\hat \sigma_{3}(\bar{\theta},\bar{k})
\end{aligned}
\end{equation}
where $\sigma_{\alpha}(\bar{\theta},\bar{k})$, $\alpha=1,2,3$ are the 3 Pauli
matrices in the basis $\left\{ e^{-\I\bar{k}\frac{\pi}{2}}\ket{\left\{ \bar{\theta}-\pi,\bar{k}\right\} },\ket{\left\{ \bar{\theta},\bar{k}\right\} }\right\} $

We show now how to obtain $\hat \sigma_{2}(\bar \theta, \bar k)$ and $\hat \Gamma_{2}^{(2)}(\bar \theta, \bar k)$ with the introduced linear and quadratic optical gates. For such, we use $\hat Z\left(\frac{1}{2}\right)=e^{-\I\frac{1}{2}\hat{\theta}}$

For $\bar{\theta}\in\left]0,\pi\right[$, we have:

\begin{align}
\hat Z\left(-\frac{1}{2}\right)\hat U\left[\frac{\pi}{2}\right]\left(\hat Z(1)+\hat Z(-1)\right)\hat U^{\dagger}\left[\frac{\pi}{2}\right]\hat Z\left(\frac{1}{2}\right)\\
=2\int_{0}^{1}d\bar{k}\intop_{0}^{\pi}d\bar{\theta}\cos\left(\bar{\theta}-\bar{k}\pi\right) & \left[e^{-\I\bar{k}\frac{\pi}{2}}\ket{\left\{ \bar{\theta}-\pi,\bar{k}\right\} }\bra{\left\{ \bar{\theta},\bar{k}\right\} }+h.c.\right]
\end{align}

Its matrix form in basis $\left\{ e^{-\I\bar{k}'\frac{\pi}{2}}\ket{\left\{ \bar{\theta}-\pi,\bar{k}\right\} },\ket{\left\{ \bar{\theta},\bar{k}\right\} }\right\} $ can be expressed as:

\begin{align}
\hat \Gamma_{2}^{(2)}= \hat Z\left(-\frac{1}{2}\right)\hat U\left[\frac{\pi}{2}\right]\left(\hat Z(1)+\hat Z(-1)\right)\hat U^{\dagger}\left[\frac{\pi}{2}\right]\hat Z\left(\frac{1}{2}\right) & =2\int_{0}^{1}d\bar{k}\intop_{0}^{\pi}d\bar{\theta}\cos\left(\bar{\theta}-\bar{k}\pi\right)\begin{pmatrix}0 & 1\\
1 & 0
\end{pmatrix}(\bar \theta, \bar k))\\
 & =2\int_{0}^{1}d\bar{k}\intop_{0}^{\pi}d\bar{\theta}\cos\left(\bar{\theta}-\bar{k}\pi\right)\hat \sigma_{2}(\bar \theta, \bar k). 
\end{align}

Using the modular variable formalism and operations, we have thus demonstrated that 
we can obtain operators that are integrals of all the $SU(2)$ generators for different $\{\bar \theta, \bar k\}$ dependent basis. In the example studied, the ``weight" function $\zeta_{\alpha}^{(2)}=\cos{(\bar \theta-\bar k \pi)}$ is the same for all $\alpha$. However, the cosine function can be replaced with any function
verifying the conditions presented in Section I
and is therefore quite general.

We conclude by connecting the ``recipe" presented here to create the $\hat \Gamma_{\alpha}^{(2)}$ operators with linear optical operations and the operators $\hat S_{\alpha}^{(2)}$ introduced in the main text. This can be done simply by defining a diagonal operator $\hat D_{\alpha}^{(d)}$ in the $\bar \theta$, $\bar k$ space. This operator not necessarily satisfies the conditions imposed in Section I. The idea is that we can express the operators $\hat \Gamma_{\alpha}^{(d)}$ as 
\begin{equation}\label{SD}
\hat \Gamma_{\alpha}^{(d)}=\hat S_{\alpha}^{(d)}\hat D_{\alpha}^{(d)}+\hat D_{\alpha}^{(d){\dagger}}\hat S_{\alpha}^{(d)\dagger}=2{\rm Re}\left [ \hat S_{\alpha}^{(d)}\hat D_{\alpha}^{(d)}\right ],
\end{equation}
an expression that is valid in the particular case of $d=2$ as well. From Section II, we see that Eq. (\ref{SD}) corresponds to the operators created in the exit port of the introduced interferometers. Moreover, expression (\ref{SD}) explicits the fact that operators $\hat \Gamma_{\alpha}^{(d)}$ are observables.

\section{Examples of $SU(3)$--type operators}

An example of solution to conditions imposed in Section I is $F(\bar \theta, \bar k)=e^{i\bar \theta-\bar k \frac{2\pi}{3}} $. This implies that $c_n=e^{\frac{2\pi n i}{3}}$ and ${\cal F}(\bar \theta, \bar k)=e^{i\bar \theta-\bar k \frac{2\pi}{3}}$. Thus, we have that $\hat \Delta$ can be written as: 
\begin{equation}\label{operator}
\hat \Delta= \int_0^{2\pi} {\rm d} \bar \theta d \bar k e^{i(\bar \theta-\bar k \frac{2\pi}{3})} (\ket{\{\bar \theta, \bar k\}} \bra{\{\bar \theta, \bar k\}}+e^{i\frac{2\pi}{3}}\ket{\{\bar \theta+\frac{2\pi}{3}, \bar k\}} \bra{\{\bar \theta+\frac{2\pi}{3}, \bar k\}}+e^{i\frac{4\pi}{3}}\ket{\{\bar \theta+\frac{4\pi}{3}, \bar k\}} \bra{\{\bar \theta+\frac{4\pi}{3}, \bar k\}}). 
\end{equation}
%\hat {\cal L}_{\alpha}=
 By defining 
\begin{eqnarray}
&&\hat \Gamma_3^{(3)} = \int_0^{2\pi/3} {\rm d} \bar \theta \cos{(\bar \theta-\bar k \frac{2\pi}{3})} \hat \lambda_3(\bar \theta,\bar k) + \sin{(\hat \theta-\bar k \frac{2\pi}{3})} \hat \lambda_8 (\bar \theta,\bar k)\nonumber  \\
&&\hat \Gamma_8^{(3)}=\int_0^{2\pi/3} {\rm d} \bar \theta \sin{(\bar \theta-\bar k \frac{2\pi}{3})}\hat \lambda_3 (\bar \theta,\bar k)- \cos{(\bar \theta-\bar k \frac{2\pi}{3})} \hat \lambda_8(\bar \theta,\bar k), 
\end{eqnarray}
where the $\hat \lambda_{\alpha}(\bar \theta, \bar k)$, with $\alpha=1,...,8$ matrices are the Gell-Mann matrices  in the subspaces $\ket{\{\bar \theta, \bar k\}},\ket{\{\bar \theta+2\pi/3, \bar k\}}, \ket{\{\bar \theta +4\pi/3, \bar k\}}$, 
we see (\ref{operator}) can be re-expressed, for $d=3$,  as:
\begin{eqnarray}
&&\hat \Delta=\hat \Gamma_{3}^{(3)}+i\hat \Gamma_{8}^{(3)}=\\
&&\int_0^{2\pi/3}\int_0^1 d \bar \theta d \bar k \left [\cos{(\bar \theta-\bar k \frac{2\pi}{3})} \hat \lambda_3(\bar \theta, \bar k) + \sin{(\hat \theta-\bar k \frac{2\pi}{3})} \hat \lambda_8 (\bar \theta, \bar k)+i(\sin{(\bar \theta-\bar k \frac{2\pi}{3})}\hat \lambda_3 (\bar \theta, \bar k) - \cos{(\bar \theta-\bar k \frac{2\pi}{3})} \bar \lambda_8 (\bar \theta, \bar k))\right ].\nonumber
\end{eqnarray}

Thus,  Eq. (4) in the main text is such that 
\begin{eqnarray}
&&\zeta_3^{(3)}(\bar \theta, \bar k)=\cos{(\bar \theta-\bar k \frac{2\pi}{3})}\nonumber \\
&&\hat \gamma_3(\bar \theta,\bar k)=\hat \lambda_3(\bar \theta,\bar k)+\tan{(\bar \theta-\bar k \frac{2\pi}{3})} \hat \lambda_8(\bar \theta,\bar k) \nonumber \\
&&\zeta_8^{(3)}(\bar \theta, \bar k)=\sin{(\bar \theta-\bar k \frac{2\pi}{3})}\nonumber \\
&&\hat \gamma_8(\bar \theta,\bar k)=\hat \lambda_3(\bar \theta, \bar k)-{\rm cotan}(\bar \theta-\bar k \frac{2\pi}{3})\hat \lambda_8(\bar \theta,\bar k)
\end{eqnarray}
The operators $\hat \gamma_3(\bar \theta,\bar k)$ and $\hat \gamma_8(\bar \theta,\bar k)$  are linearly independent for all $\bar \theta, \bar k$, i.e., in all the continuum of $3$ dimensional subspaces. Again, starting from Eq. (\ref{operator}) provides one possible solution, and others with some $\bar k$ dependency, for instance, are also possible, as mentioned in Section I. By comparing this expression to the discussion on the experimental implementation presented in Section II, we see that by setting $\hat D_{\alpha}^{(3)}=\hat \Delta$ (that corresponds to putting SLMs in each arm of the interferometer in Fig. 2) and $\hat S_{\alpha}^{(3)}=\mathbb{1}$ we have, in each exit port of the interferometer, the equivalent of the action of operators $\hat \Gamma_3^{(3)}$ and $\hat \Gamma_8^{(3)}$.

We will show now that the $6$ other non-diagonal independent generators can always be created in such subspaces by using the operations defined in Section II: $\hat U(\varphi)$, $\hat X(\Theta)$ and $\hat Z(K)$.  For such, we set $K=\pm1$ and $\varphi=\pi/2$ and consider a varying $\Theta$. The value of $\Theta$ determines $\alpha$. This can be seen by computing 
\begin{align}
X^{\dagger}\left(\Theta\right)U\left[\frac{\pi}{3}\right]\left(Z(1)+Z(-1)\right)U^{\dagger}\left[\frac{\pi}{3}\right]X\left(\Theta\right) & =\int_{0}^{1}\ud\bar{k}\intop_{0}^{\frac{2\pi}{3}}\ud\bar{\theta}\biggl(e^{-\I\bar{k}\frac{2\pi}{3}}e^{\I\left(\bar{\theta}+\Theta\right)}\begin{pmatrix}0 & 0 & e^{\I\frac{2\pi}{3}}\\
e^{-\I\frac{2\pi}{3}} & 0 & 0\\
0 & 1 & 0
\end{pmatrix}\nonumber \\
 & +e^{\I\bar{k}\frac{2\pi}{3}}e^{-\I\left(\bar{\theta}+\Theta\right)}\begin{pmatrix}0 & e^{\I\frac{2\pi}{3}} & 0\\
0 & 0 & 1\\
e^{-\I\frac{2\pi}{3}} & 0 & 0
\end{pmatrix}\biggr),
\end{align}
that can be re-expressed as 
\begin{eqnarray}\label{ND}
&&\hat {\cal C}(\Theta)=\int_{0}^{1}d\bar{k}\intop_{0}^{\frac{2\pi}{3}}d\bar{\theta}(\cos{(\bar \theta+\Theta-\frac{2\pi}{3}-\bar k \frac{2\pi}{3})}\hat \lambda_1(\bar \theta, \bar k))+\sin{(\bar \theta+\Theta-\frac{2\pi}{3}-\bar k \frac{2\pi}{3})}\hat \lambda_2(\bar \theta, \bar k))\nonumber  \\
&&-\cos{(\bar \theta+\Theta+\frac{2\pi}{3}-\bar k \frac{2\pi}{3})}\hat \lambda_4(\bar \theta, \bar k))-\sin{(\bar \theta+\Theta+\frac{2\pi}{3}-\bar k \frac{2\pi}{3})}\hat \lambda_5(\bar \theta, \bar k))\\
&&+\cos{(\bar \theta+\Theta-\bar k \frac{2\pi}{3})}\hat \lambda_6(\bar \theta, \bar k))+\sin{(\bar \theta+\Theta-\bar k \frac{2\pi}{3})}\hat \lambda_7(\bar \theta, \bar k)).\nonumber
\end{eqnarray}
From (\ref{ND}), we can chose $6$ values of $\Theta$, $\Theta_i$, $i=1,..6$, such that $[\hat {\cal C}(\Theta_i),\hat {\cal C}(\Theta_j)]\neq 0$$\  \forall \ i,j$ with $i \neq j$. This means that for every $\{\bar \theta, \bar k \}$ dependent tridimensional subspace, we can find a set of $8$ independent traceless matrices. The fact that the weight function $\zeta_{\alpha}^{(d)}$ is not the same for every $\alpha$ simply means that this set is not the same for each $\{\bar \theta, \bar k \}$ dependent subspace. 

\end{widetext}


\begin{thebibliography}{99}


\bibitem{SIMON} Simon, R., Peres-Horodecki separability criterion for continuous variable systems, {\it Phys. Rev. Lett  }  {\bf 84}, 2722 (2000). 

\bibitem{DUAN} Duan, L.-M., Giedke, G, Cirac, J. I., Zoller, P., Inseparability Criterion for Continuous Variable Systems, {\it Phys. Rev. Lett.  }{\bf 84}, 2726 (2000). 

\bibitem{TELEPORTATION} Bennett, C.H.,  Brassard, G., Cr\'epeau, C., Jozsa, R., Peres, A., Wootters, W., Teleporting an unknown quantum state via dual classical and EPR channels. {\it Phys. Rev. Lett. }{\bf 70}, 1895Ð1899 (1993). 

\bibitem{TELEPORTATIONEXP} Furusawa, A., Sorensen, J. L.,  Braunstein, S. L., Fuchs, C. A., Kimble, H. J., Polzik, E. S.,  Unconditional quantum teleportation. {\it Science} {\bf 282}, 706Ð709 (1998).

\bibitem{BELL} J. S. Bell, On the Einstein Podolsky Rosen Paradox, {\it Physics} (Long Island City, N.Y.),  {\bf 1}, 195 (1964).

\bibitem{REID} Reid, M, Drummond, P, Bowen, W, Cavalcanti, E, Lam, PK, Bachor, H, Andersen, UL,  Leuchs, G., Colloquium: The Einstein-Podolsky-Rosen paradox: From concepts to applications, {\it Rev. Mod. Phys.} {\bf  81}, 1727-1751 (2009).

\bibitem{RMPBrau} Braunstein, S. and van Lock, P., Quantum information with continuous variables, {\it Rev. Mod. Phys.} {\bf 77}, 513-577 (2005). 

\bibitem{Wedbrook} Weedbrook, C. et al., Gaussian quantum information, {\it Rev. Mod. Phys.} {\bf 84}, 621-669 (2012).

\bibitem{SHOR} Shor, P. W., Polynomial-Time Algorithms for Prime Factorization and Discrete Logarithms on a Quantum Computer, {\it SIAM J. Comput.} {\bf 26},  1484 (2007). 

\bibitem{GROVER} Grover, L. K., A Fast Quantum Mechanical Algorithm for Database Search, {\it Proceedings, 28th Annual ACM Symposium in the Theory of Computing}, 212 (1996).

\bibitem{CHSH}  Clauser, J. F., Horne, M.A. , Shimony, A., Holt, R.A., Proposed experiment to test local hidden-variable theories, {\it Phys. Rev. Lett.} {\bf 23} (15): 880 (1969). 

\bibitem{BANA} Banaszek, K. and W\'odkiewicz, K. ,Testing Non--Locality in Phase Space, {\it Phys. Rev. Lett.} {\bf  82}, 2009 (1999). 

\bibitem{pseudo}  Chen, Z.-B.,  Pan, J.-W., Hou, G.  and  Zhang, Y.-D., Maximal Violation of BellÕs Inequalities for Continuous Variable Systems, {\it Phys. Rev. Lett.}, {\bf 88}, 040406 (2002). 

\bibitem{Saleh} Yarnall, T., Abouraddy, A. F., Saleh, B. E. A, and Teich, M. C, Experimental Violation of BellÕs Inequality in Spatial-Parity Space, {\it Phys. Rev. Lett.} {\bf 99}, 170408 (2007). 

\bibitem{AHARANOV} Aharanov, Y.  Pendleton, H.  and Petersen, A.,  Modular Variables in Quantum Theory, {\it Int. J. Theor. Phys.} {\bf 2}, 212 (1969).  

\bibitem{AHARANOV2} Aharanov, Y.  and Bohm, D., Significance of Electromagnetic Potentials in the Quantum Theory, {\it Phys. Rev.} {\bf 115}, 485 (1959). 

\bibitem{RUSSOS} Gneiting, C.  and Hornberger, K. , Detecting Entanglement in Spatial Interference, {\it  Phys. Rev. Lett.} {\bf 106},
210501 (2011).

\bibitem{STEVE} Carvalho, M. A. {\it et al}, Experimental Observation of Quantum Correlations in Modular Variables, {\it Phys. Rev. A} {\bf 86}, 032332 (2012).

\bibitem{Zak1} Zak, J., Finite Translations in Solid-State Physics, {\it Phys. Rev. Lett.} {\bf 19}, 1385 (1967).

\bibitem{Zak2} Zak, J. {\it Phys. Rev.}, Dynamics of Electrons in Solids in External Fields,  {\bf 168}, 686 (1968).

\bibitem{Englert} Englert, B.-G., Lee, K.-L., Mann, A. and Revzen, M., Periodic and discrete Zak bases, {\it New J. Phys.}, {\bf 39}, 1669 (2006).

\bibitem{Comment} We stress that Eq. (3) represent a choice of basis that is made in order to better illustrate the main idea presented in this work. Other choices of basis, including global phase factors in the basis states were made throughout this Letter in order to simplify calculations. This point is discussed in \cite{SI}

\bibitem{SI} Supplementary Information available at ....

\bibitem{SAULO} Machado, S., Milman, P. and  Walborn, S. P., Interferometric scheme for direct measurement of moments of transverse
spatial variables of photons, {\it Phys. Rev. A} {\bf 87}, 058334 (2013). 

\bibitem{comment2} The denomination ``close to eigenstates" is due to the fact that eigenstates of the $\hat \Gamma_{\phi_i}^{(2)}$ are not physical, they correspond to $\delta$--like functions. However, one can defined distributions which are arbitrarily close to these eigenstates and still violate inequalities (\ref{bell}).

\bibitem{Milman07} Milman, P. , Keller, A., Charron,  E. and Atabek, O. Bell-Type Inequalities for Cold Heteronuclear Molecules,  {\it Phys. Rev. Lett.}{\bf 99}, 130405 (2007).

\bibitem{Milman10}  Milman, P. , Keller, A., Charron,  E. and Atabek, O. Molecular orientation entanglement and temporal Bell-type inequalities,  {\it Eur. Phys. J. D} {\bf 53}, 383 (2009).

\bibitem{Borges12}  Borges, C. V. {\it et al.}, Bell inequalities with continuous angular variables, {\it Phys. Rev. A} {\bf 86}, 052107 (2012).

\bibitem{COLLINS} Collins, D.,  Gisin,  N., Linden, N., Massar, S. and Popescu, S.,  Bell Inequalities for Arbitrarily High-Dimensional System, {\it Phys. Rev. Lett. } {\bf 88}, 040404 (2002). 

\bibitem{QUTRITS} Kaszlikowski, D. {\it et al}, Clauser-Horne inequality for three-state systems, {\it Phys. Rev. A} {\bf 65}, 032118 (2002). 

\bibitem{example} A detailed example of an application and of the advantages of the formalism presented here will be presented elsewhere. It consists of adapting to continuous variables systems  the Grover search algorithm. 

\bibitem{WOOTTERS} Wootters, W. K., Entanglement of formation of an arbitrary state of two qubits, {\it Phys. Rev. Lett.} {\bf 80}, 2245 (1998).

\bibitem{PERES} Peres, A., Separability Criterion for Density Matrices, {\it Phys. Rev. Lett.} {\bf 77} 1413 (1996). 

\bibitem{HORO} Horodecki, M., Horodecki, P.  and Horodecki, R., Separability of Mixed States: Necessary and Sufficient Conditions, {\it Phys. Lett. A} {\bf 223}, 1 (1996). 


\bibitem{TASCA}  Tasca, D. S.,  Gomes, R. M, Toscano, F. ,  Souto Ribeiro, P. H. and  Walborn, S. P., Continuous-variable quantum computation with spatial degrees of freedom of photons, {\it Phys. Rev. A} {\bf 83}, 052325 (2011). 




\bibitem{BVL} Lloyd, S. and Braunstein, S., Quantum computing over continuous variables, {\it Phys. Rev. Lett.} {\bf 82}, 1784 (1999). 


\end{thebibliography}
\end{document}